\documentclass[reprint,amsmath,amssymb,aps,showkeys]{revtex4-2}
\usepackage{float}
\usepackage{caption}
\usepackage{subcaption}
\usepackage{graphicx}
\usepackage{dcolumn}
\usepackage{bm}

\begin{document}

\title{DFT+U study of UO$_2$: Correct lattice parameter and electronic band-gap}

\author{Mahmoud Payami}%
\email{Email: mpayami@aeoi.org.ir}
\affiliation{School of Physics \& Accelerators, Nuclear Science and Technology Research Institute, AEOI, 
	P.~O.~Box~14395-836, Tehran, Iran
}%

\date{\today}%

\begin{abstract}
Hubbard-corrected density functional theory, denoted by DFT+U method, was developed to enable correct prediction of insulating properties for strongly-correlated electron systems. UO$_2$ is an example having O-$2p$, U-$6d$, and U-$5f$ incomplete electronic shells. Usually, researchers apply the Hubbard correction only to the localized incomplete $5f$ electrons of U atoms and succeed to predict insulating property and good geometric properties by tweaking the Hubbard-U parameter. However, it turned out that in such a way it was impossible to obtain reasonable values for both geometry and electronic band-gap at the same time. In this work, we show that it is possible to produce  good values for those properties just by applying and tuning the Hubbard corrections to all incomplete shells of O-$2p$, U-$6d$, and U-$5f$.   
\end{abstract}


\maketitle

\section{Introduction}\label{sec1}
\begin{figure}
	\centering
	\begin{subfigure}{.2\textwidth}
		\includegraphics[width=\textwidth]{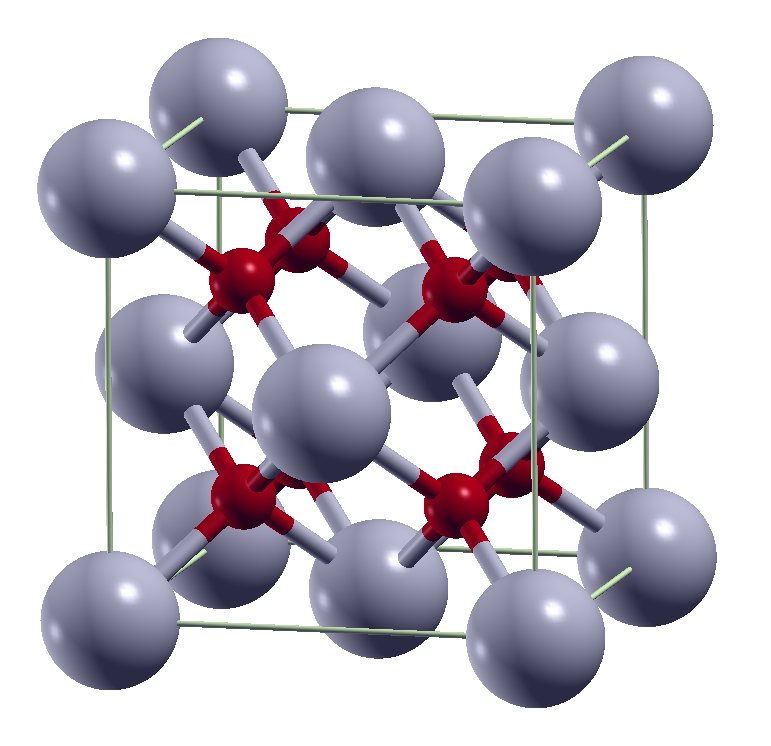}
		\caption{}
	\end{subfigure}
	\begin{subfigure}{.2\textwidth}
		\includegraphics[width=\textwidth]{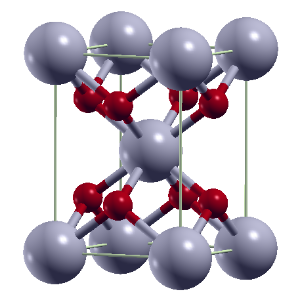}
		\caption{}
	\end{subfigure}\\
	\caption{(a)- UO$_2$ crystal structure with cubic space group $Fm\bar{3}m$ (No. 225) and lattice constant of 5.47\AA; (b)- description by a simple tetragonal crystal structure with six atoms. Gray and small red balls represent uranium and oxygen atoms, respectively.}
	\label{fig1}
\end{figure}

UO$_2$, as a common fuel for nuclear power reactors, has attracted the interests of researchers for a better theoretical description within DFT+U approachc.\cite{dorado2009dft+,payamiIJPR2022,payami2022relativistic,payami2023comparison} 
Uranium dioxide has a 3D anti-ferromagnetic (AFM) crystal structure at temperatures less than $30~$K,\cite{Amoretti,Faber} but usually a simpler 1D-AFM model is used for the description. Recent XRD experiment\cite{desgranges2017actual} has shown that UO$_2$ crystallizes with a cubic space group $Pa\bar{3}$  (No. 205). However, if the structure is modeled by a slightly different but more symmetric cubic space group $Fm\bar{3}m$ (No. 225) with experimental lattice constant of 5.47\AA, which is shown in Fig.~\ref{fig1}(a), then the structure can be represented by a simple tetragonal unit cell with 6 atoms as shown in Fig.~\ref{fig1}(b). 

Experiment has shown\cite{schoenes1978optical} that UO$_2$ is electrically an insulator with a gap of 2.10 eV.
Ordinary approximations in density-functional theory (DFT) such as local-density approximation (LDA)\cite{hohenberg1964,kohn1965self} or semi-local approximations such as generalized gradient approximation (GGA) \cite{PhysRevLett.100.136406} for the localized orbitals usually lead to incorrect metallic behavior. One workaround is to estimate the interactions of localized orbitals using the Hubbard model and add it to the DFT energy functional and then subtract the double-counting contributions from the DFT energy functional:\cite{cococcioni2005linear,himmetoglu2014hubbard,dorado2009dft+,freyss4}

\begin{equation}
	E_{DFT+U}=E_{DFT}-E_{dc}+E_{Hub}.
	\label{eq1}
\end{equation}  

The interaction term in Hubbard model, when the Hamiltonian is represented in the basis of strongly localized Wannier functions, is written as:

\begin{equation}
	E_{Hub}=U\sum_i n_{i\uparrow} n_{i\downarrow}
	\label{eq2}
\end{equation} 
where $U$ is a real number, $n_{i\sigma}$ with $\sigma=\uparrow,\downarrow$ denote the particle number operators, and $i$ specifies the lattice site ${\bf R}_i$. 
For positive values of $U$, the interaction behaves as on-site repulsion among the electrons, while on the other hand, negative values of $U$ means that there exist on-site attraction among electrons.   

In previous DFT+U calculations for UO$_2$, the on-site Hubbard correction with positive interaction parameter $U-5f$ was applied to only $5f$ electrons of uranium atoms which led to gap opening and thus correct insulating behavior. 
However, the gap size and geometric properties such as lattice constant both depend on the interaction parameter. By tuning this on-site parameter, it is possible to reproduce only one of those properties: gap size or lattice constant. As is seen from Fig.~\ref{fig2}, for the approximation used here, the correct band gap is reproduced by assuming $U_{\rm{U}-5f}$=3.2~eV while the correct lattice constant is reproduced by taking $U_{\rm{U}-5f}$=4.8~eV.   
 In this work, we have extended the Hubbard correction to cover $6d$ of U atoms as well as $2p$ of O atoms, and determine the relevant interaction parameter values that reproduce both band-gap and lattice constant of the GS in very good agreement with experiment. 

The organization of this paper is as follows. In Section~\ref{sec2}, we explain the computational details; in Section~\ref{sec3} the calculated results are presented and discussed; and finally in Section~\ref{sec4} we concludes this work. 

\section{Computational details}\label{sec2}
The DFT+U calculations are done by solution of the KS equations using the Quantum-ESPRESSO code package \cite{Giannozzi_2009,doi:10.1063/5.0005082}. Ultra-soft pseudo-potentials (USPP) are used for U and O atoms that has been generated by the {\it atomic} code, using the generation inputs (with small modifications for more desired results) from the {\it pslibrary}, \cite{DALCORSO2014337} at https://github.com/dalcorso/pslibrary.
The valence configurations of U($6s^2,\, 6p^6,\, 7s^2,\, 7p^0,\, 6d^1,\, 5f^3 $) and O($2s^2,\, 2p^4 $) were used in the generation. The relativistic effects were accounted at the level of scalar-relativistic (SR) approximation,\cite{koelling1977technique} which has been shown to give reasonable GS geometric results\cite{payami2023comparison} for $U_{\rm{U}-5f}$=4.53~eV when the Perdew-Zunger \cite{perdew1981self} (PZ) LDA approximation was used for the exchange-correlation, and the projection on to Hubbard orbitals were chosen to be atomic ones that were not orthonormalized.
The appropriate kinetic energy cutoffs for the plane-wave expansions 
were chosen as 90 and 720~Ry for the wavefunctions and densities, respectively. Also, the Methfessel-Paxton smearing method \cite{methfessel1989high} for the occupations with a width of 0.01~Ry is used for better convergency process. 
For the Brillouin-zone integrations in geometry optimizations, a $8\times 8\times 6$ grid were used. All geometries were fully optimized for total residual pressures on unit cells to within 0.5 kbar, and residual forces on atoms to within 10$^{-3}$~mRy/a.u.
The occupation matrix control (OMC)\cite {dorado2009dft+} is used to avoid metastable states. The starting magnetization for oxygen atoms are set to zero values and for U atoms they are set to $\pm0.5$ to make anti-ferromagnetic (AFM) configuration along the $z$ direction.
Since in the present work we apply Hubbard corrections to $5f$, $6d$ localized orbitals of U atoms and $2p$ orbitals of O atoms, we have occupation matrices of dimensions $7\times 7$, $5\times 5$, and $3\times 3$, respectively.
Applying the Hubbard correction for each of the orbitals U-$5f$, U-$6d$ and O-$2p$ separately one at a time showed that only the Hubbard corrections to U-$5f$ orbitals lead to metastable states and the other two are insensitive to initial occupations. 
Since in the U-atom pseudo-potential the $5f$ orbital is occupied by 3 electrons, then we have $C_3^7=35$ different ways for occupying the diagonal elements of $7\times 7$ matrix by 3 electrons: [1110000], [1101000], [1100100], $\cdots$, [0001011], [0000111].       

\section{Results and Discussions}\label{sec3}
\begin{figure}[ht]
	\centering
	\begin{subfigure}{.45\textwidth}
		\includegraphics[width=\textwidth]{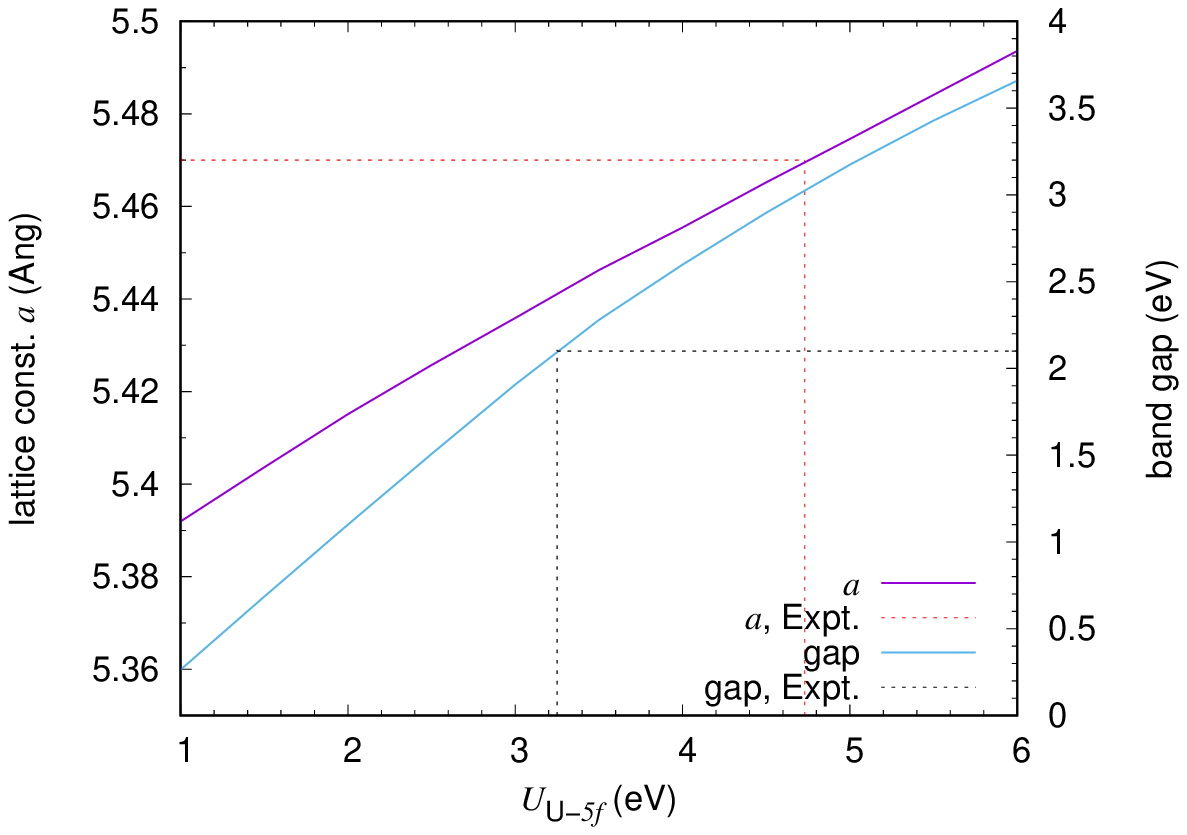}
		\caption{}
	\end{subfigure}
	\begin{subfigure}{.45\textwidth}
		\includegraphics[width=\textwidth]{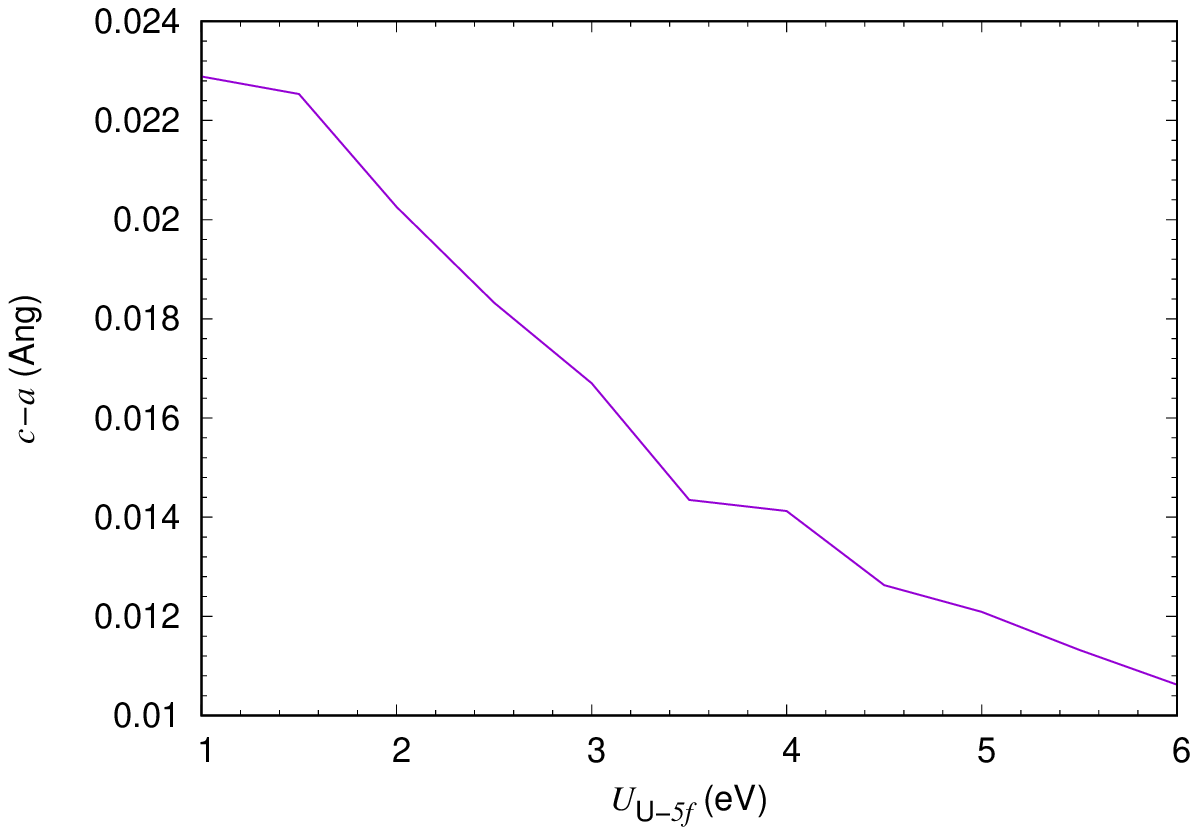}
		\caption{}
	\end{subfigure}\\
	\caption{(a)- Lattice constant $a$ in $\AA$ and band gap in eV as functions of Hubbard correction strength $U_{\rm{U}-5f}$; (b)- Deviation form cubic geometry, ($c-a$), in $\AA$ as function of Hubbard correction strength $U_{\rm{U}-5f}$ of U-atom.}
	\label{fig2}
\end{figure}

\begin{figure}[ht]
	\centering
	\includegraphics[width=0.45\textwidth]{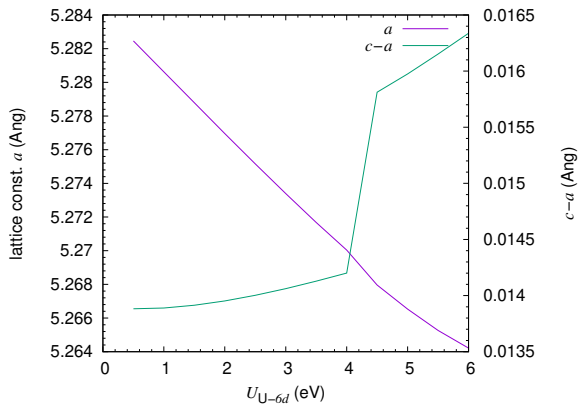}
	\caption{Lattice constant $a$ and deviation from cubic geometry ($c-a$) in $\AA$ as functions of Hubbard correction strength $U_{\rm{U}-6d}$.}
	\label{fig3}
\end{figure}

\begin{figure}[ht]
	\centering
	\includegraphics[width=0.45\textwidth]{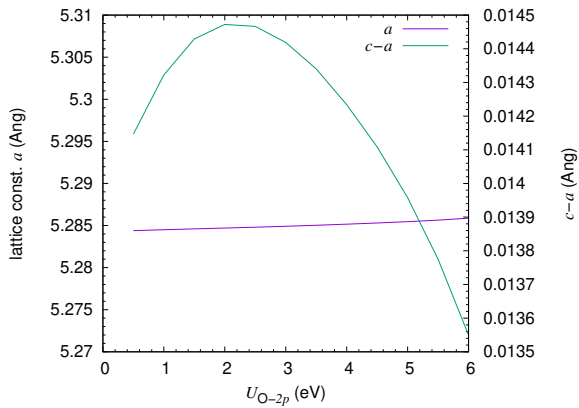}
	\caption{The same as in Fig.~\ref{fig3} but for Hubbard correction strength $U_{\rm{O}-2p}$.}
	\label{fig4}
\end{figure}

Examining Hubbard corrections for $5f$, $6d$ of U-atom and $2p$ of O-atom separately one at a time shows that only correction on $5f$ is able to open an energy gap and give reasonable lattice constant. The situation is shown in Figs.~\ref{fig2}-\ref{fig4}.     

Inspecting Fig.~\ref{fig2}(a) it is seen that the value for experimental lattice constant is reproduced around $U_{\rm{U}-5f}\sim 4.8$~eV while the correct band gap is reproduced by assuming $U_{\rm{U}-5f}$=3.2~eV. This implies that with only one correction parameter (i.e., $U_{\rm{U}-5f}$) one fails to reproduce reasonable values for both lattice constant and the band gap at the same time. 

In Fig.~\ref{fig2}(b), the deviation from cubic geometry ($c-a$) is shown to be very small, $\sim 0.01\AA$, so that modeling the system by 1D-AFM (instead of 3D-AFM) does not cause any significant error in this study.     
On the other hand, Figs.~\ref{fig3} and \ref{fig4} show that the value of lattice constant is relatively insensitive to the values $U_{\rm{U}-6d}$ and $U_{\rm{O}-2p}$. These results hint that one should do fine-tuning of $U_{\rm{U}-5f}$ around the value of 4.0~eV. 
In addition, similar to the result in Fig.~\ref{fig2}(b), the deviations from cubic geometries in Figs.~\ref{fig3}-\ref{fig4} are negligible.  

\begin{figure}[ht]
	\centering
	\includegraphics[width=0.45\textwidth]{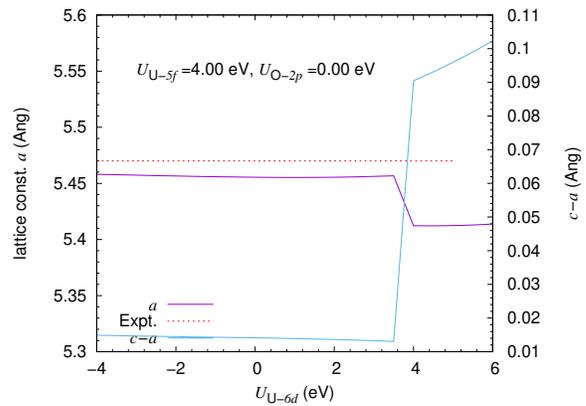}
	\caption{The same as in Fig.~\ref{fig3} as functions of Hubbard correction strength $U_{\rm{U}-6d}$ for fixed value of $U_{\rm{U}-5f}$=4.00~eV.}
	\label{fig5}
\end{figure}

In the next step, we apply Hubbard corrections to two orbitals at a time. In above it was shown that the correct lattice constants were reproduced by applying the correction to only U-$5f$ with the strength of $\sim 4.00$~eV. So, we consider the correction to U-$5f$ with strength 4.00~eV as the main correction and add that for U-$6d$ as a background one with different values. The result is shown in Fig.~\ref{fig5}. As is seen from Fig.~\ref{fig5}, adding the background correction for U-$6d$ almost does not change the lattice constant for $U_{\rm{U}-6d} < 3.0$~eV and so we ignore the background correction for U-$6d$.  

\begin{figure}[ht]
	\centering
	\includegraphics[width=0.45\textwidth]{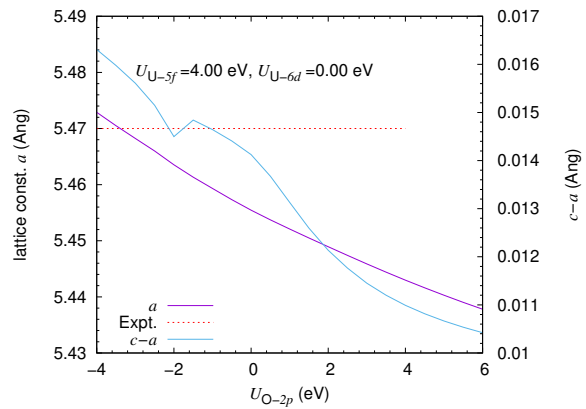}
	\caption{The same as in Fig.~\ref{fig5} as functions of Hubbard correction strength $U_{\rm{O}-2p}$ for fixed value of $U_{\rm{U}-5f}$=4.00~eV.}
	\label{fig6}
\end{figure}

We now concentrate on adding the background correction of O-$2p$ orbitals. As is seen from Fig.~\ref{fig6}, in contrast to the case of U-$6d$, here the background correction to O-$2p$ orbitals significantly modifies the results attained by the correction on U-$5f$. That is, in order to maintain the reasonable value for the lattice constant, one should use negative values for the Hubbard correction parameter for O-$2p$ orbitals, meaning that the background correction is as an on-site attraction. To summarize, Fig.~\ref{fig6} indicates that the combination of two simultaneous corrections with $U_{\rm{U}-5f}=4.0$~eV and $U_{\rm{O}-2p}=-3.00$~eV revives the reasonable value for the lattice constant. But now we expect that the electronic band gap is changed from the value 2.91~eV, obtained if only U-$5f$ correction was applied.\cite{payami2023comparison}
Fig.~\ref{fig6} also indicates that the deviation from cubic geometry is still acceptable.

To have a closer inspection on the effect of negative values for $U_{\rm{O}-2p}$, we have calculated the GS lattice constants and electronic band gaps for different values of $U_{\rm{U}-5f}$, keeping $U_{\rm{O}-2p}$ fixed at three values of -3.00, -3.50, and -4.00~eV. The results are presented in Table~\ref{tab1}.

\begin{table}[ht]
	\caption{\label{tab1}%
		 Lattice constants in $\AA$ and $E_{gap}$ in eV as functions of $U_{\rm{U}-5f}$ and fixed values of background correction $U_{\rm{O}-2p}$=-3.00, -3.50, and -4.00~eV.}
	\begin{ruledtabular}
		\begin{tabular}{lccc}
			$U_{\rm{O}}$ (eV)& $U_{\rm{U}}$ (eV) & $a$ ($c$) ($\AA$)   &  $E_{gap}$ (eV)  \\ \colrule 
-3.00 &  3.00 &  5.4477 (5.4667) &   1.7651 \\
      &  3.10 &  5.4498 (5.4684) &   1.8397 \\
      &  3.20 &  5.4518 (5.4701) &   1.9096 \\
      &  3.30 &  5.45391(5.4718) &   1.9706 \\
      &  3.40 &  5.4559 (5.4734) &   2.0308 \\
      &  3.50 &  5.4579 (5.4751) &   2.0900 \\
      &  3.60 &  5.4599 (5.4767) &   2.1484 \\
      &  3.70 &  5.4619 (5.4784) &   2.2059 \\
      &  3.80 &  5.4644 (5.4795) &   2.2623 \\
      &  3.90 &  5.4666 (5.4817) &   2.3180  \\
      &  4.00 &  5.4682 (5.4838) &   2.3731 \\ \hline
-3.50 &  3.00 &  5.4499 (5.4693) &   1.7387 \\
      &  3.10 &  5.4520 (5.4710) &   1.8125 \\
      &  3.20 &  5.4541 (5.4727) &   1.8789 \\
      &  3.30 &  5.4561 (5.4744) &   1.9387 \\
      &  3.40 &  5.4581 (5.4759) &   1.9977 \\
      &  3.50 &  5.4601 (5.4776) &   2.0556 \\
      &  3.60 &  5.4622 (5.4794) &   2.1129 \\
      &  3.70 &  5.4642 (5.4812) &   2.1691 \\
      &  3.80 &  5.4665 (5.4830) &   2.2245 \\
      &  3.90 &  5.4685 (5.4847) &   2.2789 \\
      &  4.00 &  5.4705 (5.4864) &   2.3320 \\ \hline
-4.00 &  3.00 &  5.4522 (5.4720) &   1.7112 \\
      &  3.10 &  5.4543 (5.4737) &   1.7842 \\
      &  3.20 &  5.4564 (5.4754) &   1.8472 \\
      &  3.30 &  5.4584 (5.4769) &   1.9057 \\
      &  3.40 &  5.4604 (5.4786) &   1.9636 \\
      &  3.50 &  5.4626 (5.4805) &   2.0206 \\
      &  3.60 &  5.4648 (5.4823) &   2.0765 \\
      &  3.70 &  5.4669 (5.4841) &   2.1315 \\
      &  3.80 &  5.4689 (5.4858) &   2.1855 \\
      &  3.90 &  5.4708 (5.4874) &   2.2383 \\
      &  4.00 &  5.4728 (5.4891) &   2.2902 \\
		\end{tabular}
	\end{ruledtabular} 
\end{table}
 
\begin{figure}[ht]
	\centering
	\includegraphics[width=0.45\textwidth]{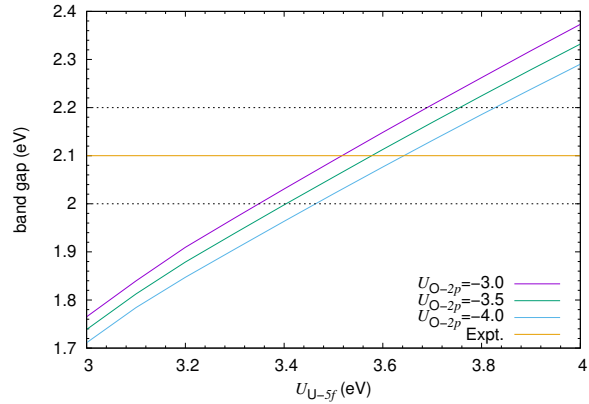}
	\caption{Variation of band gap as function of $U_{\rm{U}-5f}$ for different fixed values of $U_{\rm{O}-2p}$. It is seen that, in this example, to obtain each desired values of $E_{gap}$=2.00, 2.10, and 2.20~eV one has three choices for the Hubbard strength combinations.}
	\label{fig7}
\end{figure}

In order to estimate the proper combinations of Hubbard strengths for $U_{\rm{U}-5f}$ and $U_{\rm{O}-2p}$ for a desired value of band gap (2.00, 2.10, 2.20~eV), we have plotted the data of Table~\ref{tab1} in Fig.~\ref{fig7}.

As we see from Fig.~\ref{fig7}, since here we have chosen three fixed values for $U_{\rm{O}-2p}$, there exist three different combinations of the Hubbard strengths $U_{\rm{U}-5f}$ and $U_{\rm{O}-2p}$ for each desired value of band gap. To verify the validity of this guess, we have calculated the GS properties for nine combinations of Hubbard strengths hinted by plots of Fig.~\ref{fig7} and presented the results in Table~\ref{tab2}.  
 
\begin{table}[ht]
	\caption{\label{tab2}%
	Hubbard parameters $U_{\rm{U}}$, $U_{\rm{O}}$, in eV, needed to be used to obtain a desired band gap $\tilde{E}_{gap}$ along with the resulted lattice constants $a\;(c)$ and band gap $E_{gap}$.	}
	\begin{ruledtabular}
		\begin{tabular}{lccc}
			$\tilde{E}_{gap}$ (eV) & $U_{\rm{U}}$, $U_{\rm{O}}$ (eV)&  $a$ ($c$) ($\AA$)   & $E_{gap}$ (eV)  \\ \colrule 
			2.00 &  3.40, -3.00 &  5.4560 (5.4735) &   2.03 \\
			     &  3.45, -3.50 &  5.4592 (5.4768) &   2.03 \\
			     &  3.48, -4.00 &  5.4622 (5.4802) &   2.01 \\ \hline
			2.10 &  3.50, -3.00 &  5.4579 (5.4751) &   2.09 \\
			     &  3.60, -3.50 &  5.4623 (5.4795) &   2.11 \\
			     &  3.65, -4.00 &  5.4658 (5.4832) &   2.10 \\ \hline
			2.20 &  3.70, -3.00 &  5.4619 (5.4784) &   2.21 \\
			     &  3.78, -3.50 &  5.4661 (5.4827) &   2.21 \\
			     &  3.84, -4.00 &  5.4697 (5.4865) &   2.21 \\
		\end{tabular}
	\end{ruledtabular} 
\end{table}

From the data in Table~\ref{tab2}, we see that applying simultaneous Hubbard on-site corrections on the U-$5f$ and O-$2p$ orbitals it is possible to tune both lattice constant and band gap to their experimental values. 
 
\section{Conclusions}\label{sec4}

In previous theoretical studies of UO$_2$ crystal, in order to predict correct insulating behavior, researchers used Hubbard corrections for the U-$5f$ localized orbitals in the DFT+U approach. It was already shown that depending on what XC functional is used and whether the Hubbard orbitals were orthonormalized or not, for a given Hubbard-$U$ parameter (say 4.0~eV) different results were obtained for equilibrium lattice constant and the KS band gap. None of those results were satisfactory in predicting simultaneous reasonable values for the lattice constant and the size of band gap. In this work, employing LDA-PZ scheme for the XC energy functional, we have shown that applying the on-site Hubbard corrections simultaneously to U-$5f$ and O-$2p$ orbitals one can choose certain values to obtain results for both the lattice constant and energy band gap of the ground state in good agreement with experiment.

\section*{Acknowledgement}
This work is part of research program in School of Physics and Accelerators, NSTRI, AEOI.  

\section*{Data availability }
The raw or processed data required to reproduce these results can be shared with anybody interested upon 
sending an email to M. Payami.

\bibliography{payami-OMC-Gap_Tuning-1401.11.22}

\end{document}